Static solitons, Lorentz invariance, and a new perspective on the integrability of the Sine Gordon equation in (1+2) dimensions

Yair Zarmi
Jacob Blaustein Institutes for Desert Research
Ben-Gurion University of the Negev
Midreshet Ben-Gurion, 84990, Israel

Abstract
Contrary to the common understanding, the Sine-Gordon equation in (1+2) dimensions does have $N$-soliton solutions for any $N$. The Hirota algorithm allows for the construction of static $N$-soliton solutions (i.e., solutions that do not depend on time) of that equation for any $N$. Lorentz transforming the static solutions yields $N$-soliton solutions in any moving frame. They are scalar functions under Lorentz transformations. In an $N$-soliton solution in a moving frame, ($N$-2) of the (1+2)-dimensional momentum vectors of the solitons are linear combinations of the two remaining vectors.

The Sine-Gordon (SG) equation in (1+1) dimensions is integrable [1, 2]. However, its (1+2)-dimensional version,

$$\frac{\partial^2 u}{\partial t^2} - \frac{\partial^2 u}{\partial x^2} - \frac{\partial^2 u}{\partial y^2} + \sin u = 0 \ , \tag{1}$$

is not integrable in the Painlevé sense [3-6], or through inverse scattering [7]. Its one- and two-soliton solutions are found through an obvious extension of the solution algorithm [8] for soliton solutions of the (1+1)-dimensional SG equation. A three-soliton solution does not exist in general. It can be constructed, provided the soliton wave-number vectors obey a constraint [9]. Modified versions that are integrable have been proposed in the literature [10, 11]. In this note, it is argued that Eq. (1) is integrable in a different sense.

The solitons and anti-solitons show up in the current density, given by:

$$\vec{J} = \vec{\nabla} u(t,x,y) \qquad \left(\vec{\nabla} \equiv (\partial_x, \partial_y)\right) \ . \tag{2}$$

The algorithm starts with the Hirota transformation [8, 9]:

$$u(t,x,y) = 4 \arctan\left[\frac{g(t,x,y)}{f(t,x,y)}\right] \ . \tag{2}$$

The single-soliton solution is obtained for

$$g(t,x,y) = e^{\theta(t,x,y;p_x,p_y)} \qquad f(t,x,y) = 1 \ , \tag{3}$$

with

$$\theta(t,x,y;p_x,p_y) = p_\mu x^\mu + \delta(q) = p_0 t - p_x x - p_y y + \delta(q) \ , \qquad p_0 = \sqrt{p_x^2 + p_y^2 - 1} \ . \tag{4}$$

In all solution discussed, the free phase shifts, $\delta(q)$, will be set to zero.

To prepare for the discussion of the properties of solutions under Lorentz transformations, the traditional wave numbers are written as $p_\mu$, a momentum-vector in the (1+2)-dimensional Minkowski space; light velocity is $c = 1$. Also, note that the Hirota algorithm stipulates that $p_\mu$ is a space-like vector. Namely, $p_\mu p^\mu = -1$. The solution generated by Eq. (3) is given by:

$$u(t,x,y) = 4\tan^{-1}\left[e^{\theta(t,x,y;p_x,p_y)}\right] . \tag{5}$$

The current density vector is proportional to a single-soliton profile, and is given by:

$$\vec{J} = \frac{-2}{\cosh\left[\theta(t,x,y;p_x,p_y)\right]}\{p_x,p_y\} . \tag{6}$$

In Eq. (6), $\{p_x, p_y\}$ denotes a two-dimensional vector.

The two-soliton solution is obtained with

$$g(t,x,y) = e^{\theta\left(t,x,y;p_x^{(1)},p_y^{(1)}\right)} + e^{\theta\left(t,x,y;p_x^{(2)},p_y^{(2)}\right)} , \qquad f(t,x,y) = 1 + a_{1,2}\, e^{\theta\left(t,x,y;p_x^{(1)},p_y^{(1)}\right)} e^{\theta\left(t,x,y;p_x^{(2)},p_y^{(2)}\right)} . \tag{7}$$

In Eq. (7), the phase $\theta$, is defined by Eq. (4), and

$$a_{n,m} = \frac{\left(p_x^{(n)} - p_x^{(m)}\right)^2 + \left(p_y^{(n)} - p_y^{(m)}\right)^2 - \left(p_0^{(n)} - p_0^{(m)}\right)^2}{\left(p_x^{(n)} + p_x^{(m)}\right)^2 + \left(p_y^{(n)} + p_y^{(m)}\right)^2 - \left(p_0^{(n)} + p_0^{(m)}\right)^2} . \tag{8}$$

There is a limitation on the wave numbers. To avoid degeneration into a single-soliton solution or the existence of singularities in the current density, $a_{1,2}$ must be positive.

The three-soliton solution is obtained with

$$\begin{aligned}g(t,x,y) &= e^{\theta\left(t,x,y;p_x^{(1)},p_y^{(1)}\right)} + e^{\theta\left(t,x,y;p_x^{(2)},p_y^{(2)}\right)} + e^{\theta\left(t,x,y;p_x^{(3)},p_y^{(3)}\right)} \\ &\quad + a_{1,2}\,a_{1,3}\,a_{2,3}\, e^{\theta\left(t,x,y;p_x^{(1)},p_y^{(1)}\right)} e^{\theta\left(t,x,y;p_x^{(2)},p_y^{(2)}\right)} e^{\theta\left(t,x,y;p_x^{(3)},p_y^{(3)}\right)} \\ f(t,x,y) &= 1 + \sum_{1 \le i < j \le 3} a_{i,j}\, e^{\theta\left(t,x,y;p_x^{(i)},p_y^{(i)}\right)} e^{\theta\left(t,x,y;p_x^{(j)},p_y^{(j)}\right)}\end{aligned} \tag{9}$$

Here, again, $a_{n,m}$ must all be positive. However, there is an additional limitation. Equation (9) yields a solution only if the three momentum-vectors, $p^{(i)}$, ($i$ = 1, 2, 3), are linearly dependent [9]. No construction of $N$-soliton solutions with $N > 4$ has been presented in the literature.

The constraint on the three momentum vectors in the three-soliton solution is the key point to overcoming the situation just described. To this end, let us return to the Hirota algorithm [7, 8] and construct static single-soliton solutions (solutions that do not depend on time). Let us begin with the single-soliton case. A static solution exists, and the momentum vector, $p$, of Eq. (4) obtains a simpler form, for which a special notation is used:

$$p \rightarrow q = \{0, \cos\varphi, \sin\varphi\} \ . \tag{10}$$

The vector $q$ is also space-like, with norm $q_\mu q^\mu = -1$. Hence, a Lorentz transformation exists, which transforms $q$ into the vector $p$ of Eq. (4). In a (1+2) dimensional Minkowski space, a Lorentz boost depends on two parameters, $v_x$ and $v_y$. (The velocity of light is $c$ = 1.) The transformation of $q$ into $p$ is, therefore, a single-parameter family. The final conclusion is that any single-soliton solution of the (1+2) dimensional SG equation is associated via a Lorentz transformation with a static single-soliton solution. One might, therefore, first construct a static single-soliton solution, and then obtain a moving solution by applying the appropriate Lorentz boost *both* to the momentum vector, $p$, and to the coordinates, $\{t, x, y\}$.

The two-soliton solution of Eq. (7) depends on two momentum vectors, $p^{(i)}$ ($i$ = 1, 2). There is one Lorentz transformation, which transforms these two vectors into two space-like vectors, $q^{(i)}$, with the structure given in Eq. (10), but with two different angles, $\varphi_i$. The boost parameters are:

$$v_x = \frac{p_0^{(2)} p_y^{(1)} - p_0^{(1)} p_y^{(2)}}{p_x^{(2)} p_y^{(1)} - p_x^{(1)} p_y^{(2)}} , \qquad v_y = \frac{p_0^{(1)} p_x^{(2)} - p_0^{(2)} p_x^{(2)}}{p_x^{(2)} p_y^{(1)} - p_x^{(1)} p_y^{(2)}} \ . \tag{11}$$

Hence, as in the case of the single-soliton solution, any two-soliton solution is associated via a Lorentz transformation with a static two-soliton solution. Direct computation of the static two-soliton solution yields a simple expression for the coefficients $a_{i,j}$ of Eq. (8):

$$a_{i,j} = \tan\left[\frac{1}{2}(\varphi_i - \varphi_j)\right]^2 . \tag{12}$$

One might, therefore, generate a static two-soliton solution, and then obtain a moving solution by applying the appropriate boost *both* to the momentum vectors, $q^{(i)}$, and to the coordinates, $\{t, x, y\}$.

The three-soliton solution of Eq. (9) depends on three momentum vectors, $p^{(i)}$ ($i = 1, 2, 3$). If they are not in a plane, i.e., they are linearly independent, then there is no Lorentz boost that will transform all three vectors into three space-like vectors, $q^{(i)}$ ($i = 1, 2, 3$), with the structure given in Eq. (10). The reason is that $q^{(i)}$ lie in a plane in the (1+2) dimensional space, hence they are linearly dependent. Thus, if a three-soliton solution exists in a moving frame, it cannot be associated with a static three-soliton solution, unless its three momentum vectors are linearly dependent. This is precisely where the result of Hirota [9] enters. He found that *there is no three-soliton solution in a moving frame unless the three momentum vectors are linearly dependent*. Hence, the conclusion arrived at, concerning single- and two-soliton solutions, applies here as well. There are solutions of three moving solitons. The easiest way to find them is to generate a static three-soliton solution, and then obtain a moving solution by applying the desired boost to *both* the momentum vectors, $p^{(i)}$, and to the coordinates, $\{t, x, y\}$. Since in the static solutions the three momentum vectors, $q^{(i)}$, are linearly dependent, so are the three momentum vectors, $p^{(i)}$, of the three moving solitons.

The Hirota algorithm allows for the construction of $N$ static soliton solutions for any $N$. For each static soliton, the momentum vector is given by Eq. (4), and the coefficients, $a_{i,j}$ – by Eq. (12). (A simple MATHEMATICA program has produced static $N$-soliton solutions for $N = 1$–5.) Any Lo-rentz boost in the (1+2) dimensional space will generate $N$ moving soliton solutions, in which the $N$ momentum vectors, $p^{(i)}$, are linearly dependent. Only two of the vectors are independent.

This reopens the question of the meaning of integrability. In the traditional sense, (Painlevé, or inverse scattering), Eq. (1) is not integrable. Still, it has an infinite family of soliton solutions in any Lorentz frame, that are obtained from the static solutions by the appropriate Lorentz transformation, that is to be applied to *both* the momentum and position vectors. This transformation leaves the scalar product $p_\mu x^\mu$ in Eq. (4) invariant, and makes the soliton solutions of Eq. (1) into scalar functions under Lorentz transformations.

Acknowledgment

The author thanks M. Boiti and F. Pempinelli for encouraging comments.